\documentclass[preprint, 12pt]{aastex} 
    
\begin{document} 

\title{On the role of stochastic Fermi acceleration in setting the dissipation scale of turbulence in the interstellar medium}
\author {Robert Selkowitz and Eric G. Blackman}
\affil{Dept. of Physics and Astronomy, and Laboratory for Laser Energetics, 
University of Rochester, Rochester NY, 14627}

\begin{abstract}

We consider the dissipation by Fermi acceleration of magnetosonic turbulence in the Reynolds Layer of the interstellar medium.
The scale in the cascade at which electron acceleration via stochastic Fermi acceleration (STFA) becomes comparable to further
cascade of the turbulence defines the inner scale. For any magnetic
turbulent spectra equal to or shallower than Goldreich-Sridhar 
this turns out to be $\ge 10^{12}$cm, which is 
much larger than the shortest length scales observed in radio 
scintillation measurements.  
While STFA for such spectra then contradict models of scintillation which 
appeal directly to an extended, continuous turbulent cascade,
 such a separation of scales is consistent with the recent 
work of \citet{Boldyrev2} and \citet{Boldyrev3} suggesting
that interstellar scintillation 
may result from the passage of radio waves through the galactic distribution of thin ionized boundary surfaces of HII regions, 
rather than density variations from cascading turbulence.
The presence of STFA dissipation also provides a mechanism for the non-ionizing heat source observed
in the Reynolds Layer of the interstellar medium \citep{Reynolds}.  STFA accommodates the 
proper heating power, and the input energy is rapidly thermalized within the low density Reynolds layer plasma. 

\end{abstract}
\bigskip

\section{Introduction}

Radio scintillation has long been associated with interstellar turbulence (e.g. Minter and Spangler 1997)).  A major requirement 
of turbulence-based scintillation models is that the inner scale of the cascade is comparable to the smallest scale of 
the scintillation.  An alternative model has been proposed \citep{Boldyrev1, Boldyrev2, Boldyrev3} 
in which the scintillation instead results from index of refraction variations at  the photoionized surfaces of a 
non-Gaussian distribution of either warm ISM regions or HII clouds.  We find that Stochastic Fermi Acceleration (STFA) 
is a sufficiently efficient process in the Reynolds Layer that it imposes a cutoff at a scale too large to be consistent 
with radio scintillation, supporting the non-Gaussian cloud model.  

Observations of line ratios by \citet{Reynolds} using the Wisconsin H-alpha Mapper (WHAM) indicate that
the Reynolds layer of the Milky Way interstellar medium (ISM) cannot be heated solely by photoionization.  
Variation with galactic latitude, $|z|$, of $[SII]/H\alpha$ 
and $[NII]/H\alpha$ both are found to be consistent with increases in electron temperature, $T_e$, but not in the ionization fractions
of either species.  Additionally, studies of external galaxies NGC891, NGC 4631, and NGC 3079 \citep{reynolds2, Reynolds3} show 
similar evidence of supplemental, non-photoionizing heat sources in the Reynolds layer, the region of low density, strongly ionized 
gas located $\backsim1$kpc away from the galactic midplane.  These studies supplement the  $[SII]/H\alpha$ 
and $[NII]/H\alpha$ ratios with analysis of  $[OIII]/H_{\beta}$ and $[OII]/H_{\alpha}$, which also are consistent with 
non-ionizing heating.  The heating rate in the 
Milky Way is consistent with being 
either proportional to electron density, $n_e$, or  
density independent.  The heat source is empirically 
found to have a power input of $\epsilon = G_1 n_e$, where $G_1 \backsim 10^{-25}$erg s$^{-1}$, or $\epsilon = G_2$, where $G_2 \backsim
10^{-27}$ erg s$^{-1}$ cm$^{-3}$. This allows the supplemental heat source to dominate at high $|z|$, where $n_e$ is low and 
photoionization heating to dominate at low $|z|$ where $n_e$ is high.
A number of potential mechanisms are presented by \citet{Reynolds} as supplemental heat sources, 
including photoelectric grain heating, Coulomb collisions
with cosmic rays, magnetic field reconnection, and dissipation of supperbubble driven magnetic turbulence.  

We consider the role of STFA in the dissipation of interstellar turbulence, looking primarily for the inner scale of the cascade.  
ISM turbulence is driven principally by superbubbles, the large blast shells carved out when the members of an OB association
reach the supernova stage in a short time period.  
STFA occurs when electrons traveling in along a magnetic field line encounter 
moving compressions of the local magnetic field and are reflected. 
While STFA drains turbulence primarily into 
ions for thermal pressure dominated plasmas, electrons are the primary
energy recipient when the ion speed 
is sub-Alfv\'enic (i.e. \citet{LaRosa, b99,Selkblack}).
The latter case is relevant for the present study because  
the Reynolds layer seems to be a weakly magnetically dominated 
plasma \citep{ms96}.  

%[{\bf NEW STUFF AND COMMENTS FOR ROB BELOW}]

%In this context it should be noted that the turbulence in the Reynolds
%layer can be supermagnetosonic. 
%This raises the possibility that some fraction of
%the  energy from the turbulence can be dissipated into shocks
%as well as the general question of the spectrum of
%the turbulent cascade. We will discuss this further in the next
%section.

%Supermagnetosonic turbulence tends
%to have a steeper spectrum than incompressible turbulence for this
%reason (e.g. \cite{vos03}). 

%(Rob, see \cite{vos03} their table 2 and conclusions and intro. note that
%for their quantities 11/3 corresponds to kolomogorov)
%Note the steeper spectra, maybe we should address in
%section 2:in particular have an equation in which 
%the spectral index can be input by hand.
%Also, in GS i believe the magnetic and velocity spectra are matched
%but in real turbulence the two are different--generally
%i think we want mag spec.
%I think we should allow reader to put in their favorite
%choice of power law in section two.

%\cite{lg01} considered compressible turbulence primarily in
%the super-Alfv\'enic regime, but briefly discuss the
%sub-Alfv\'enic regime on very small scales. In our present 
%calculation, those scales are not reached.

%[{\bf END COMMENTS}]

The energy
change, $\delta E$ from a single STFA mirroring is \citep{Fermi, a81, Longair, Selkblack}

\begin{equation}
\frac{2E}{c^2}(v^2_A \pm v_e v_A),
\end{equation}
where $E$ is the initial energy of the reflected particle, $c$ is the speed of light, $v_A$ is the Alfv\'en speed, and $v_e$ is the
electron speed.  Typical Reynolds layer temperatures are $0.6 < T_4 < 1.2$ where $T_4 = T_e/10^4$K.  This corresponds to an energy in the range 
$0.5<E<1.1$eV.  Thus $\delta E$ is significantly below ionization energies, which are tens of eV, and even if the reflection rate
is high, STFA of protons is a non-ionizing process in the turbulent Reynolds layer of the ISM.

In section 2 we show that STFA provides the correct power and that the energy input is quickly thermalized, consistent with
the conditions imposed on the heating source.  We then determine the truncation scale for the turbulent cascade.   In section 3, 
we discuss the implications of the inferred STFA dissipation scale for models of interstellar scintillation, 
finding that STFA truncates the cascade at a long length scale, inconsistent with turbulence models of 
radio scintillation.  We conclude in section 4.

\section{Stochastic acceleration in the turbulent Reynolds layer}

The power available for electron heating in the MHD turbulent cascade, $\epsilon_T$, can be estimated as

\begin{equation}
\epsilon_T = n_e m_p \frac{v^3_A}{L},
\label{epsilon}
\end{equation}
where $n_e$ is the electron number density, $v_A$ is the local Alfv\'en speed, and $L$ is the outer scale of the interstellar
turbulence inertial range.  From \citet{Spangler}: $v_A = 2.3 \times 10^6$ cm s$^{-1}$ and $L=10^{19}$cm.   \cite{ms96} infer this value for $L$ 
by analysis of emission measure and rotation measure structure functions.
From \citet{Reynolds}: $n_e = 0.28$cm$^{-3}$
at galactic latitude $|z| = 1$kpc.  $\epsilon_T = 5 \times 10^{-26}$ erg s$^{-1}$ cm$^{-3}$, consistent with the heating rate 
called for in \citet{Reynolds} and calculations of superbubble injected power \citep{MacLowrev, Elmegreen}.

\citet{Selkblack} examined STFA in the $v_A \gg c_s$ regime with applications to solar flares, and concluded that the post-acceleration 
spectrum is dominated by the rate of escape from the acceleration region.  The post-escape spectrum is important in flares 
because the observed x-ray emission is generated via Bremsstrahlung as electrons escape the acceleration region and encounter the 
dense chromospheric plasma. 
In the turbulent Reynolds layer, the acceleration region is effectively infinite in extent.   Instead of observing emission by 
escaped electrons, the observed emission comes from electrons which remain in the Reynolds layer.  In this case we are concerned with 
the spectrum of electrons still confined within the acceleration region.  Regardless, the acceleration rate found in \citet{Selkblack}
remains valid

\begin{equation}
\left(\frac {dE}{dt}\right)_S = \left<\delta E\right> R,
\end{equation}
where
\begin{equation}
\left<\delta E\right> = 4 m_e v^2_A,
\end{equation}
is the average energy per reflection and 

\begin{equation}
R = \frac{v_d}{2L} \left(\frac{\lambda_s}{L}\right)^2, 
\end{equation}
is the rate of reflections, such that
\begin{equation}
\left(\frac {dE}{dt}\right)_S = \frac{2}{L} m_e v^2_A v_d {\left(\frac{\lambda_s}{L}\right)}^{-(1-1/a)},
\label{dEdt}
\end{equation}
and $\lambda_s$ is the turbulent inner scale 
(associated with the parallel
component of the magnetic field gradient for anisotropic turbulence), $v_d$ is the effective relative speed of the electrons and compressions  
within the plasma, $m_e$ is the electron mass, and $a$ is the spectral index of the turbulent cascade
\begin{equation}   
\frac{\lambda}{L} = \left(\frac{\delta B}{B}\right)^a.
\label{GS}
\end{equation}
It is assumed throughout our analysis that at length scales comparable to L the strength of the magnetic fluctuations, $\delta B$ is comparable to
that of the mean magnetic field, $B$.  The magnetic spectrum of turbulence in the ISM is 
difficult to measure.
Between $10^{19}{\rm pc}$ and $0.03$pc the spectrum 
appears to be close to Kolmogorov $(a=3)$ (and flatter on larger scales)
(\cite{hfm04,ms96}). There is no direct measure 
of the magnetic spectrum 
on scales below $0.03$pc. 

For the range of magnetic spectra $2 \le a \le \infty$ the 
result from our calculations to follow, that the cascade truncation scale 
well exceeds the smallest scintillation scale,  
will not change. However, we  
first consider a Goldreich-Sridhar (hereafter GS) turbulent spectrum with $a=2$ \citep{GS} (see also \cite{m98}).
Like \cite{m98}, 
GS turbulence is fully anisotropic, and is based on an incompressible
cascade. It involves a more rapid cascade in the direction perpendicular to the local mean field than parallel to it.   
Turbulence becomes less and less compressible on smaller
scales, and since the Reynolds layer seems to be 
modestly magnetically dominated (\cite{ms96}), GS is plausible on small enough
scales: The GS spectrum more closely arises in compressible and incompressible 
simulations when an initially relative strong field is imposed
(e.g. \cite{mg01},\cite{cho2002a},\cite{vos03},\cite{hb04b}), i.e. when 
the initial ratio of thermal to magnetic pressure, $\beta <1$
and the velocity fluctuations are of order the initial field.
Other values for $a$ should not be ruled out however, 
because the magnetic turbulent spectrum 
for galactic ISM conditions is not a solved theoretical problem.
In particular, when an initially weak field is imposed, driven incompressible
simulations show that the magnetic spectral index may be flatter,
and flatter than the velocity spectra.
Although it may approach $5/3$ for magnetic Prantdl number
of order unity (\cite{hb03}) there may be a trend toward
further flattening at larger magnetic Prandtl numbers 
(\cite{sch02,mcm04,sch04,hb04}).
The magnetic spectrum is steeper in the presence of kinetic
helicity (\cite{maronblackman02}). The compressible driven simulations 
of  \cite{vos03} show that the stronger the initial
field, the closer the magnetic and velocity spectra match,
and the flatter the magnetic spectra the weaker the initial field.
The role of the magnetic Prandtl number is hard to assess in
\cite{vos03} since no explicit viscosity or resistivity is used.

Proceeding with $a=2$,
we note that for this case,
\citet{Selkblack} find
that the parallel cascade law is relevant to STFA, and $\lambda$ in eq. (\ref{GS}) corresponds to the parallel cascade law. 
When electrons are able to freely stream from one STFA scattering site to the next without deflection by pitch angle scattering 
(the free streaming limit) $v_d$ is equal to the electron speed $v_e$ and, 

\begin{equation}
\left(\frac {dE}{dt}\right)_{f} = \frac{2}{L} m_e v^2_A v_e \left(\frac{\lambda_{f}}{L}\right)^{-1/2}.
\label{dEdtfree}
\end{equation}
In order to determine the truncation scale of the cascade, we impose the balance condition that the turbulent power, $\epsilon_T$ is equal to
the STFA acceleration rate.  Setting the STFA rate equal to the total power selects a single value of $\lambda_f$, which 
is an effective upper limit on the truncation scale.  We use the subscript $f$ to denote the free streaming limit.

In the limit of very strong pitch angle scattering, where the scattering length scale $\lambda_p \ll \lambda_{f}$, $\lambda_p$ can 
be considered the electron mean free path. 
Electrons are effectively trapped in regions smaller than $\lambda_f$, executing a random walk with 
small drift speed.  They encounter magnetic 
compressions only as the compressions stream past, significantly reducing the acceleration rate below that of Eq \ref{dEdtfree}.  
The power balance condition between STFA and the cascade power is not met, 
and the cascade will proceed to scales shorter than $\lambda_f$.  The cascade cannot continue farther than a length scale comparable to $\lambda_p$.  
At this point, the system passes back into a free-streaming regime, with a substantially shorter $\lambda$ than is required to 
meet the power balance condition.  The turbulence then drains rapidly.

From the above two paragraphs, we conclude that the cascade 
would be truncated at a scale no smaller than the minimum of $\lambda_p$ and $\lambda_f$.
To assess the  minimum scale of dissipation for the Reynolds layer, we assume that the pitch angle scattering is 
dominated by electron-electron Coulomb scattering. 
\citet{Spitzer} finds the Coulomb self-collision time, $t_c$, to be

\begin{equation}
t_c = \frac{0.266}{n_e} \frac{T^{3/2}_e}{\ln \Lambda},
\end{equation}
where $T_e$ is the electron temperature, and $\ln\Lambda \backsim 25$ is determined by the effective long range cutoff of the Coulomb
force in a plasma.  The electron speed is given by $v_e = \sqrt{2kT/m_e}$, and thus

\begin{equation}
\lambda_p = v_e t_c = \sqrt{\frac{2kT}{m_e}} \frac{0.266}{n_e} \frac{T_e^{3/2}}{\ln\Lambda} = 2 \times 10^{13} T^2_4,
\end{equation}
where $T_4 = T_e/10^4$.For the range of observed temperatures in \citet{Reynolds}, $0.6 < T_4 < 1.2$, the electron mean free path falls in
the range $8 \times 10^{12}$cm $\lambda_p < 3 \times 10^{13}$cm.  Taking the lower limit of this range 
fixes the lower bound of the cascade truncation scale.  

In the free streaming limit we can find $\lambda_{f}$ by setting  $\epsilon_T = n_e (\frac{dE}{dt})_{f}$,

\begin{equation}
\lambda_{f} = \left(\frac{2m_e}{m_p}\right)^2 \left(\frac{v_e}{v_A}\right)^2 L = 6 \times 10^{14}  T^{-1}_4.
\label{lambdafree}
\end{equation}
For the observed temperature range of $0.6 < T_4 < 1.2$, $10^{15}$cm$ > \lambda_{f} > 5 \times 10^{14}$cm.  This sets an upper bound on the 
cascade truncation scale.

It should be noted that the free streaming
approximation is invalid in the Reynolds layer as $\lambda_{f}/\lambda_p = 100$ at $T_4 = 6$.  There are, on average 
100 pitch angle scattering events per electron per encounter with a compression.  Given the low fraction of encounters  which result
in a reflection, $F$,  electrons cannot stream freely from one reflection site to the next.  
The quantity $F$ is determined by the pitch angle condition for reflection

\begin{equation}
\cos^2{\theta_{min}} < \frac{\delta B}{B},
\end{equation}
and the turbulent cascade law 
(e.g. eq. (\ref{GS}) for GS turbulence). 
All electrons with $\theta > \theta_{min}$ reflect.  Assuming pitch angle isotropy, we estimate

\begin{equation}
 F = \frac{1}{4 \pi} \int_{\theta_{min}}^{\pi} 2 \pi \sin({\theta}) d\theta =  \frac{1}{2} \left(\lambda_{f}/L\right)^{1/4},
\end{equation}
where we have integrated over all angles greater than $\theta_{min}$ to find the fraction of phase space which satisfies the 
pitch angle condition for reflection, and we have used (\ref{GS}) for
the last equality.
Electrons which have too large a component of their momentum in the direction 
parallel to the field are not stopped by the compressions, and pass through them.
The acceleration rate must be retarded significantly.  We have bounded the turbulent dissipation scale 
$8 \times 10^{12}$cm$<\lambda_s<10^{15}$cm for GS turbulence for a Goldreich-Sridhar, $a=2$, cascade.

In order to produce a dissipation scale consistent with the smallest scales
observed in scintillation measurements, the cascade must truncate at the inner scale predicted by scintillation models.
Stated values of this scale vary: $3.5 \times 10^6$cm in \citet{moran}, $3 \times 10^7$cm in \citet{Molnar}, 
and $3 \times 10^{10}$cm in \citet{rickett}.  Rewriting eq \ref{lambdafree} for an arbitrary value of $a$, we have
\begin{equation}
\lambda_{f} = \left(\frac{2m_e}{m_p}\right)^{1-1/a} \left(\frac{v_e}{v_A}\right)^{1-1/a}. 
\end{equation}
In order to produce a value $\lambda_f=10^{10}$cm, the cascade must have a very steep spectrum, with $a < 1.3$. The other predictions
require even steeper spectra.  Such spectra are inconsistent with 
simulations of interstellar turbulence, as well as the inferred spectral indices of scintillation models, 
which place the lower limit near a Kolmogorov spectrum,
$a = 3$ \citep{hfm04, Spangler}.  

Assuming the Reynolds layer is weakly magnetically
dominated \citep{ms96} implies that 
we have studied dissipation of 
fast magnetosonic mode turbulence.
On the other hand, \citet{lg01} argue that 
for thermally dominated plasmas, $\beta > 1$, the turbulent density  fluctuations are dominated by the slow and entropy mode, and 
that the fast mode is decoupled from the slow, entropy, and Alfv\'en modes.  
Of the slow and entropy modes, only the slow mode is important for STFA.
While the fast mode could be dominant for STFA in the Reynolds layer, the acceleration by slow modes does achieve 
the same qualitative result, albeit at a shorter dissipation scale.   
In the Reynolds layer, where $\beta \backsim 0.1$, the slow mode velocity is given roughly by the sound speed $c_s \backsim 0.2 v_A$. 
If both modes are present with equal energy density, the fast mode is more efficient for STFA than 
the slow mode.  To apply eq. (\ref{dEdt}) to STFA of slow modes in a plasma with $\beta < 1$ requires the replacement of $v_A$ with 
$c_s$, such that

\begin{equation}
\left(\frac {dE}{dt}\right)_S = \frac{2}{L} m_e c^2_s v_d {\left(\frac{\lambda_s}{L}\right)}^{-(1-1/a)}.
\end{equation}
For free streaming electrons, there is no change in $v_d = v_e$ and the dissipation scale is only decreased by a factor of 
$(c_s/v_A)^2 = 0.04$.  For trapped electrons, the mean free path continues to place a lower bound on the dissipation.  
Despite the moderate decrease in $\lambda_s$ in a slow mode dominated cascade, it remains well above
the scintillation length scale.

In addition to providing the correct power input and a physically plausible dissipative scale, STFA must supply energy in a 
manner consistent with heating, as 
opposed to non-thermal acceleration.  \citet{Selkblack} demonstrated that, in the limit of no self-interaction which is appropriate 
for solar flare applications above a few keV, the electron energy spectrum within the acceleration region
is shaped by two competing effects: a bulk shifting to higher energy, and a diffusive spreading of the spectrum.  The resulting
spectrum can be quasi-thermal, even when electrons are not truly sharing energy.  However, in the case of ISM heating, where Coulomb 
self-scattering of electrons is presumed to be the dominant source of pitch angle scattering, STFA is accompanied by rapid 
thermalization of the electron population.
Consider the acceleration time, $\tau_{STFA}$

\begin{equation}
\tau_{STFA} = \frac{E}{\left(\frac{dE}{dt}\right)_s} = 3 \times 10^{12} T_4 s,
\end{equation}
and the thermalization time, $\tau_{eq}$, \citep{Spitzer}

\begin{equation}
\tau_{eq} = 6 \times 10^5 T^{3/2}_4 s,
\end{equation}
where we have taken the thermalization time to be the two species equilibration time, with both species being electrons at a single 
temperature, $T_4$.  When $\tau_{eq} < \tau_{STFA}$, as is the case for the Reynolds layer, then the energy input via STFA is shared 
among electrons rapidly; the energy spectrum is thermal.  The electron-proton thermalization time is greater than  
$\tau_{eq}$ by a factor of the mass proton to electron ratio $m_p/m_e = 1836$, which is nevertheless shorter than the acceleration time.  
The protons (and heavy ions) remain in equilibrium with the electrons as well.

\section{Implications of the turbulent cutoff scale for models of interstellar scintillation}

The dominant power source for interstellar turbulence is likely supernovae and superbubble shells (\citet{Elmegreen, MacLowrev}). 
%Superbubbles are the regions 
%carved out by the collective blast wave which forms when many members of an OB association undergo supernova explosions over a short 
%time period.  
Superbubbles are powerful and frequent enough to periodically pass through, and thus reseed turbulence in, 
the entire galactic disk. 
If the efficiency for converting a single supernova's mechanical luminosity of $10^{51}$erg to turbulent energy is 
$\eta_{SN} = 0.1$, the rate of supernova events in the Milky Way is $R_{SN}= 50$yr$^{-1}$, then the total turbulent driving from 
supernovae and superbubbles is estimated to be $3 \times 10^{-26}$erg s$^{-1}$ (reviews by \citep {MacLowrev, Elmegreen} and 
references within).  This is consistent with the estimated damping rate $\epsilon_T$ from eq (1) as well as the required 
supplemental heating rate \citep{Reynolds}.

Although we have considered a GS cascade law 
because it is inherently anisotropic, and incorporates the magnetic field into 
the formalism directly,  we emphasize again that 
the GS cascade, with $a=2$ magnetic spectral index 
lies at one end of a range of possible cascade power laws 
(e.g. \cite{GS,vos03}).
%There is evidence both observationally \citep{Spang, Spangler} and numerically \citep{vos03} that the magnetic spectral index of ISM turbulence
%may correspond to a Kolmogorov cascade law, $a=3$, and the maximal value found in the simulations of \citet{vos03} is $a=7/2$.
As discussed, varying $a$ over a wide range does not substantially alter the results of our analysis.   

Likewise, it is not certain which magnetosonic wave mode, fast or slow, is dominant in the Reynolds layer.  \citet{lg01} argues for 
slow modes to dominate in a high $\beta$ plasma.  
Fast mode turbulence may play a stronger role in  
$\beta \backsim 0.1$ diffuse Reynolds layer plasma.  Because the fast mode is a more efficient STFA accelerator, due primarily to its
higher phase speed, we emphasized STFA on fast modes,
but it was shown in section
2 that the major qualitative result of our study holds for either wave mode: 
STFA is a sufficient source of heating, and the dissipation
scale of MHD turbulence is longer than the short scales observed in radio scintillation measurements.

The correspondence between the turbulent driving power and the inferred supplemental heating rate of \citet{Reynolds} is suggestive of a 
connection.  This connection would be more strongly verifiable if the turbulent dissipation scale could be  linked to direct observations of the Reynolds layer.  
This has not yet been done.
\citet{Spang} and \citet{Spangler} considered that  
the radio scintillation 
observations  (e.g. Armstrong, Rickett, Spangler 1995) 
may be consistent with 
scattering by localized turbulent density fluctuations across the inertial range of Kolmogorov turbulence, but they further suggest that the
``fluctiferous'' medium is either HII region envelopes, or the higher density portions of the warm ionized medium (WIM).
Although possible turbulent dissipation scales (like the Larmor radius)
of the scintillating region 
correspond well with the smallest scales of radio scintillation, 
no scintillation
measurements are analyzed in lines of sight through the low density Reynolds layer.  \citet{Spangler} instead assume that either the 
Larmor radius or the ion inertial length is the key dissipative scale in the Reynolds layer, just as in denser regions of the ISM, and use this assumption to 
calculate dissipation and heating rates.  

We find that the STFA turbulent dissipation scale is instead likely far larger ($\lambda_s \ge 10^{13}$cm) in the Reynolds layer, for 
GS turbulence.  
The STFA dissipation scale does drop to the observed scintillation scale in the case of a very steep turbulent spectrum at small 
scales, with index $a \leq 1.3$.  Interpretations of the scintillation as turbulence driven, such as \citet{Spangler}, indicate
a spectrum no steeper than Kolmogorov, $a=3$.    
Although STFA is seemingly
at odds with radio scintillation measurements, it may not be if the source
of the scintillation is not a continuous turbulent cascade as is commonly assumed.
In particular, recent studies have found that interstellar radio 
scintillation  is in fact consistent with a non-Gaussian distribution of stellar ionization shells 
\citet{Boldyrev1, Boldyrev2, Boldyrev3}.  Local index 
of refraction variations at the thin photoionized surfaces of molecular clouds are distributed randomly throughout the interstellar
medium.  Radio waves are deflected through small angles at these fronts, and adjacent rays can be deflected through differing paths.  
A detailed model employing Levy statistics provides an excellent fit to the temporal spreading of pulses from distant pulsars, which
traditional cascade scintillation models have had difficulty explaining.
\citet{Boldyrev2} and \citet{Boldyrev3} conclude that the density fluctuations are more likely to result from randomly distributed
thin ionization shells in the ISM than from global turbulence.  As a result, the scintillation would no longer  provide a 
constraint on the inner scale of the turbulent cascade.
Turbulence driven on very large ($\ge$ 10 pc) scales could
damp on scales larger than the smallest scintillation scales
 and dissipation of the turbulence by STFA  
would not conflict with scintillation measurements.

The lower bound of the dissipation scale found above  violates an assumption used in the approach of  \citet{Spangler}.
Given that the measured supplemental heating rate in the low density ISM can be well explained by 
turbulent dissipation, one may infer that turbulence is driven and dissipated throughout the ISM.  
The density is higher and temperature lower in the lower $|z|$, and thus $\lambda_p$ is shorter, in the  portions of the WIM where 
the scintillation may occur.  STFA is then even more closely tied to the lower bound for $\lambda_s$ set by 
Coulomb scattering at low $|z|$.  If STFA is important, it
 would then be unlikely that turbulence input by superbubbles
cascades to sufficiently
small scales in any region of the interstellar medium to account for the scintillation observations.  This gives credence to the 
argument of \citet{Boldyrev1, Boldyrev2, Boldyrev3}, which does not rely on cascading turbulence but concludes that the statistical 
spatial distributions of the ionized boundaries of molecular clouds better explains scintillation than do models based on 
turbulence dominated density fluctuations.

\section{Conclusion}

Dissipation of turbulence almost certainly plays a role in the heating of the interstellar medium.  The turbulent 
energy supply from supernovae is sufficient to provide the supplemental heating source required by the observations of
\citet{Reynolds, reynolds2, Reynolds3}.  We have shown that STFA can act as the damping mechanism in the Reynolds layer of the Milky 
Way, truncating the turbulent cascade at length scales no shorter than $8 \times 10^{13}$cm for a cascade no steeper than Goldreich-Sridhar.
This truncation scale is too large to be consistent with turbulence-based models of interstellar scintillation.  
Instead, models like those of \citet{Boldyrev1, Boldyrev2, Boldyrev3}, which do not rely on the cascade, are required.

{}

\end{document}